\documentclass[10pt]{article}
\usepackage[letterpaper]{geometry}
\usepackage{indentfirst}
\usepackage{graphicx}
\usepackage{hicss}
\usepackage{times}
\usepackage[none]{hyphenat}
\usepackage{url}
\usepackage{latexsym}
\usepackage{listings}
\title{Dual-Technique Privacy \& Security Analysis for E-Commerce Websites Through Automated and Manual Implementation}

\author{Urvashi Kishnani \\
 University of Denver \\
 {\underline{urvashi.kishnani@du.edu}} \\  \And
 Sanchari Das \\
 University of Denver \\
 {\underline{sanchari.das@du.edu} } \\ }

\date{}

\begin{document}
\maketitle
\begin{abstract}
As e-commerce continues to expand, the urgency for stronger privacy and security measures becomes increasingly critical, particularly on platforms frequented by younger users who are often less aware of potential risks. In our analysis of $90$ US-based e-commerce websites, we employed a dual-technique approach, combining automated tools with manual evaluations. Tools like CookieServe and PrivacyCheck revealed that $38.5\%$ of the websites deployed over $50$ cookies per session, many of which were categorized as unnecessary or unclear in function, posing significant risks to users' Personally Identifiable Information (PII). Our manual assessment further uncovered critical gaps in standard security practices, including the absence of mandatory multi-factor authentication (MFA) and breach notification protocols. Additionally, we observed inadequate input validation, which compromises the integrity of user data and transactions. Based on these findings, we recommend targeted improvements to privacy policies, enhanced transparency in cookie usage, and the implementation of stronger authentication protocols. These measures are essential for ensuring compliance with CCPA and COPPA, thereby fostering more secure online environments, particularly for younger users.
\end{abstract}

\subsubsection*{Keywords:}

E-Commerce Websites, Security, Privacy

\section{Introduction}
By $2027$, global digital transactions are projected to reach an astonishing US\textdollar$9.04$ trillion~\footnote{\url{https://www.statista.com/outlook/dmo/fintech/digital-payments/worldwide}}, underscoring the critical need for robust privacy and security measures in e-commerce. These transactions frequently involve the exchange of sensitive PII such as names, addresses, email accounts, and credit card details\cite{trautman2015commerce,ackerman1999privacy,kishnani2022privacy}. This high volume of PII exchange presents a lucrative target for cybercriminals, amplifying the risks of data breaches\cite{broadhurst2017cybercrime}. The repercussions of such breaches are profound, as seen in the $2013$ Yahoo data breach that compromised nearly $3$ billion accounts, highlighting the potential for massive financial losses and irreparable damage to brand reputation~\footnote{\url{https://www.csoonline.com/article/2130877/the-biggest-data-breaches-of-the-21st-century.html}}. In $2018$, online businesses were the target in $32.4\%$ of all successful cyber-attacks, emphasizing the vulnerability of e-commerce platforms~\footnote{\url{https://www.getastra.com/blog/knowledge-base/ecommerce-security/}}.

Given the critical importance of e-commerce platforms in today's digital landscape, privacy policies, cookie practices, and payment security must be central to protecting user data and informing users about data handling practices\cite{kretschmer2021cookie,wheeler2022user}. However, these essential areas often receive less attention, leading to potential vulnerabilities, especially on platforms frequented by younger users\cite{montgomery2017ensuring,kishnani2023assessing}. To explore this, our research is guided by the following key research questions:

\noindent \textbf{RQ1:} How do current privacy policies of e-commerce platforms align with privacy protection regulations?

\noindent \textbf{RQ2:} What cookies are most commonly used by e-commerce sites, and do these platforms transparently communicate the purposes of these cookies to users?

\noindent \textbf{RQ3:} How secure are e-commerce platforms for end-users during financial transactions, and do they provide workflows for secure transaction?

To address these, we conducted an analysis of $90$ e-commerce platforms, employing the PrivacyCheck V3 tool\cite{nokhbeh2022privacycheck,zaeem2021privacycheck} to assess $20$ key aspects of privacy policies, focusing on their transparency and compliance with regulations such as CCPA and COPPA~(\cite{R95}). Additionally, we used CookieServe\cite{tay2023ensuring} to categorize cookies based on their primary function—necessary, analytical, or advertising—thus providing insights into how these websites manage user data~\footnote{\url{https://www.cookieserve.com/}}. We complemented these automated tools with a manual evaluation of critical security features, including authentication methods, digital certificate use, input validation, and overall user trust.

\noindent \textbf{C1:} Our dual-technique evaluation of $90$ e-commerce websites reveals significant privacy and security gaps, such as the lack of clear breach notification protocols and the absence of mandatory two-factor authentication.

\noindent \textbf{C2:} We identified instances where e-commerce platforms fall short of meeting privacy standards set by laws like CCPA and COPPA. These regulations, inspired by GDPR, offer a comprehensive framework for evaluating the adequacy of website privacy policies.

\noindent \textbf{C3:} To our knowledge, this is the first study to focus on cookie-centric analysis and privacy policy assessment for e-commerce platforms frequently visited by younger users, including those related to games and toys.

\section{Related Works} \label{related_work}
E-commerce security involves securing critical components such as data transport protocols, web servers, clients, and network operating systems, which are essential for preventing unauthorized access and fostering trust in online platforms~(\cite{R49,R56,R10,R52,R63,R69}). A persistent challenge in e-commerce is credit card fraud\cite{cherif2023credit,basin2023inducing}, necessitating improved security for servers, transactions, and payment processes~(\cite{R47,das2020risk}). E-commerce security covers multiple dimensions—integrity, non-repudiation, authenticity, confidentiality, privacy, and availability—each vital for protecting digital assets~(\cite{R26,R43,das2019privacy}). Educating stakeholders, including customers, financial institutions\cite{das2021organizational}, and merchants, about information security is crucial for enhancing e-commerce security~(\cite{josang2007survey,kishnani2022privacy}). 

Regarding privacy, cookies are widely used by e-commerce websites to track consumer behavior and personalize content~(\cite{schiefermair2020effects}). The regulatory environment necessitates the use of cookie notices to comply with legal requirements, which significantly impact consumer perceptions of privacy and trust~(\cite{brazhnik2013cookies,fu2020cookie}). Our analysis of cookie types and distribution in our dataset sheds light on these practices. Additionally, adherence to privacy policies is critical, as research shows that clear privacy notices can build consumer trust and influence purchase decisions~(\cite{broeder2020culture}). Further studies emphasize the need for privacy policies that are more comprehensible, either through visual aids or concise formatting~(\cite{R103, R106}). Our study assesses the compliance of e-commerce websites with privacy guidelines and their effectiveness in safeguarding user privacy.

Trust is fundamental to e-commerce success, influenced by factors like user satisfaction and security measures~(\cite{R23,R67,R73,das2020user,kishnani2023towards}). Research indicates that trust, security, privacy, and risk perceptions are interlinked on e-commerce platforms~(\cite{R54,R101,R102}). The sustained growth of e-commerce depends on user trust in secure payment processes and the platform's overall credibility~(\cite{R6,R55,gopavaram2019iotmarketplace}). Previous work has focused on identifying security vulnerabilities and implementing defensive strategies~(\cite{R17,R45,R92}). Authentication mechanisms are another crucial aspect of e-commerce security, affecting user perceptions of website safety~(\cite{R29,R61,R66,das2019evaluating}). Users often prioritize aspects like interface design, content quality, and site reliability, but the robustness of authentication systems—whether traditional or biometric—plays a significant role in shaping their security perceptions~(\cite{R5,R14,R72,das2020mfa}). This study builds on previous research to offer a comprehensive analysis of e-commerce security and privacy, integrating both user and business perspectives.

\section{Methodology} \label{methodology}

\subsection{Resource Gathering}
Our study focuses on e-commerce websites selling physical goods, particularly those frequented by US-based younger users under the age of $16$, to assess adherence to CCPA (for US-based users) and COPPA (for younger users). We began by selecting appropriate website categories based on Moiseev's classification~(\cite{moiseev2016classification}), identifying $15$ distinct categories. We excluded \lq Auto Products\rq and \lq Medical Goods\rq due to their appeal to older demographics, retaining $13$ categories. Considering the rise in online sales of Cannabis~(\cite{caputi2018online}) and Weapons~(\cite{prasad2000challenges}), these categories were added. Additionally, \lq Games and Toys\rq were included to represent younger users, resulting in a total of $16$ categories. Table~\ref{tab:categories} provides details of the distribution of websites across these categories.

\begin{table}[t]
\caption{Website Categories with Number (\#) and Percentage (\%) of Websites in Each Category}
\label{tab:categories}
\centering
\begin{tabular}{lrr}
\hline
\textbf{Category}                                 & \textbf{\#} & \textbf{\%}    \\ \hline
Clothing and Footwear                    & $19$ & $21.11$ \\
Games and Toys                           & $14$ & $15.56$ \\
Sport Equipment and Hobbies              & $9$  & $10.00$    \\
General Stores                           & $9$  & $10.00$    \\
Technical and Industrial Equipment       & $6$  & $6.67$  \\
Appliances and Electronics               & $6$  & $6.67$  \\
Household Goods                          & $6$  & $6.67$  \\
Souvenirs, Presents                      & $5$  & $5.56$  \\
Food and Kindred Products                & $4$  & $4.44$  \\
Media (Books, Concert Tickets) & $3$  & $3.33$  \\
Health and Beauty Products               & $2$  & $2.22$  \\
Furniture                                & $2$  & $2.22$  \\
Cannabis                                & $2$  & $2.22$  \\
Weapons                                  & $1$  & $1.11$  \\
Pet Supplies                             & $1$  & $1.11$  \\
Jewellery and Clocks                     & $1$  & $1.11$   \\ \hline
\textbf{Total}                           & \textbf{90}  & \textbf{100} \\ \hline
\end{tabular}
\end{table}

Websites were selected based on their accessibility in the US and availability on the Clearnet, excluding those hosted on the Darknet. To be included, a website had to sell only goods and provide an online payment option. Websites offering guest checkout were also included. We excluded websites that did not ship to or sell to US-based customers. Some websites without US-based domains were included if they accepted US-issued credit cards and provided services to US residents. A diverse selection of websites was chosen from areas such as general online stores, sports equipment, clothing, electronics, and toys, to ensure a representative sample and minimize sampling bias. Ultimately, $90$ e-commerce websites were analyzed, a sample size supported by prior research~(\cite{law2006study}). We reached out to the organizations three months prior to submitting this work to report the identified vulnerabilities and share our analysis. After discussions with senior researchers and following the disclosure protocols in prior work\cite{sun2014detecting}, we decided not to disclose the specific names of the websites in this paper due to ethical considerations and the lack of response from some of the websites.

\subsection{Website Analysis}

We conducted the website analysis in three primary steps: privacy policy, cookies, and manual interactions.

\subsubsection{Privacy Policy Analysis}

For the privacy policy analysis, we employed PrivacyCheck v3~(\cite{R95}), a browser extension available on the Chrome Web Store~\footnote{\url{https://chrome.google.com/webstore/category/extensions}}. PrivacyCheck utilizes data mining to automatically summarize the text of privacy policies through a machine learning algorithm, presenting an \lq at-a-glance\rq\ format. The tool evaluates privacy policies from two perspectives: user control and GDPR compliance. PrivacyCheck assesses $20$ different subcategories, $10$ under user control and $10$ under GDPR compliance, as shown in Table~\ref{table:privacy_scores}. 

Each subcategory is scored from $0$ to $10$, with the overall scores for user control and GDPR compliance represented as a percentage of the total possible score for each category. User control subcategories include Email Address Security, PII Security, SSN Security, Targeted Advertising, Location Tracking, COPPA Compliance, Law Enforcement, Privacy Policy Opt-Out, Data Control, and Data Aggregation. GDPR subcategories cover Between Site Transfer, Company Location, Right to be Forgotten, Data Retention Notification, Reject Usage of PII, Under $16$ Protection, Data Encryption, Data Processing Consent, Data Protection Principles, and Breach Notification. While our study focuses on US-based websites subject to the CCPA, we used PrivacyCheck because of the significant overlap between CCPA and GDPR, in terms of personal data definitions, user rights, and business obligations~(\cite{wong2023privacy}). However, for aspects where CCPA does not clearly overlap with GDPR, we provided additional manual analysis to align with CCPA requirements. 

For standards that do not clearly overlap between GDPR and CCPA, we offer the following clarifications: (i) For \textbf{Data Retention Notification}, CCPA does not explicitly require businesses to inform consumers if they retain personal information for legal purposes after a deletion request, although it emphasizes transparency by requiring businesses to inform consumers about their data collection and processing practices in privacy notices. (ii) For \textbf{Data Encryption}, while CCPA does not mandate disclosing encryption practices, it is considered a best practice for informing users about data security measures. The CCPA's primary focus is on data privacy rights rather than specific security measures. (iii) For \textbf{Data Processing Consent}, CCPA requires businesses to inform consumers about their data processing practices in privacy notices. Although explicit informed consent is not always required under CCPA, transparency about data collection and processing is crucial for compliance. (iv) For \textbf{Data Protection Principles}, CCPA does not require implementing data protection principles by design and default, but it encourages privacy-by-design principles to enhance data protection. We used these guidelines to evaluate privacy policies under CCPA, supplementing PrivacyCheck's analysis where necessary.

\subsubsection{Cookies Analysis}

We used a tool called Cookieserve~\footnote{\url{https://www.cookieserve.com/}} as used in prior work~(\cite{barneslgbtq}). Cookieserve is a free online tool, hosted on a website and powered by Cookieyes and available to the public~(\cite{carneiro2021platform}). The tool allows the user to enter a website's URL and gives as output not only the number of cookies but their general function, description of use, original domain and period of their use~(\cite{tay2023ensuring}). This tool scans the provided URL and detects cookies established by the website on that particular page. Subsequently, it classifies these cookies into distinct groups according to their properties: (i) \textbf{Necessary}: that are essential for a website's basic functionality while ensuring security; (ii) \textbf{Analytical}: that capture user's interaction with the website and provide metrics like number of visitors, bounce rate, traffic source, etc.; (iii) \textbf{Functional}: that perform functionalities like sharing content of the website to social media platforms, collect feedbacks, and provide third-party features; (iv) \textbf{Performance}: that are used to understand and analyze the key performance indexes of the website which helps in delivering a better user experience for the visitors; (v) \textbf{Advertisement}: that are used to provide visitors with relevant ads, marketing campaigns, and to collect information for customized ads; and (vi) \textbf{Other}: uncategorized cookies. 

\subsubsection{Manual Website Analysis}

Here we focused on key security features: Website Security, Customer Confidence, Authentication, Payment Security, and Input Validation. We began by assessing the website's overall security, starting with an initial impression by examining the URL after loading the website in a browser. We checked whether the website was using Secure Socket Layer (SSL) and Hypertext Transfer Protocol Secure (HTTPS). By clicking on the padlock icon next to the URL, we expanded the security information menu to verify the validity of the website's digital certificate. We then evaluated customer confidence by determining whether the website provided a platform for customer reviews and ratings. This feature, whether through star ratings or written text reviews, is crucial for assessing user trust and confidence in the website. 

For authentication, we assessed the availability of basic authentication methods, such as username and password, and checked whether multi-factor authentication (MFA) was offered. We also noted if the website permitted guest checkouts without requiring account creation. Regarding payment security, we checked whether the website used recognized payment processors and e-commerce platforms, and whether it displayed any associations with Financial Institutions (FIs) or required Electronic Signatures (ES) to verify billing information. Finally, in evaluating input validation, we deliberately entered incorrect information into the billing/shipping address fields and credit/debit card details (e.g., invalid zip codes or card numbers) to assess the website's ability to prevent fraudulent transactions and ensure data accuracy. 

\section{Results and Discussions} \label{results}
\subsection{Tool-based Analysis}
\begin{figure}[tbp]
    \includegraphics[width=\linewidth]{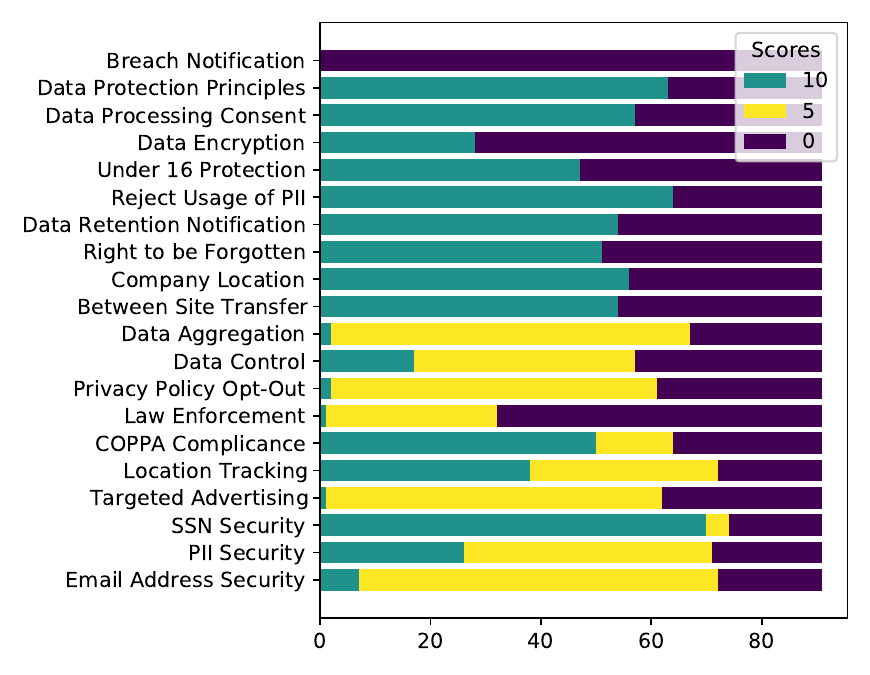}
    \caption{PrivacyCheck Score distribution against percentage of websites for privacy policies}
    \label{figure:privacy_policies}
\end{figure}

\textbf{Privacy Policy Analysis (RQ1):}
We found that the average user control score across the evaluated e-commerce websites was $53.4\%$, indicating that many sites offer only limited control over user data. In contrast, compliance with broader privacy policy frameworks, such as CCPA and GDPR, was somewhat higher, with an average score of $60.78\%$, but approximately $40\%$ compliance still missing. However, significant gaps remain, particularly in areas like breach notification and data encryption, where the policies were often either vague or completely silent. The detailed scores for the user control and policy framework categories, along with their respective subcategories, are summarized in Table~\ref{table:privacy_scores}. Each subcategory is scored on a scale from $0$ to $10$, reflecting various aspects of privacy compliance. Figure~\ref{figure:privacy_policies} illustrates the distribution of these scores, showing the percentage of websites that received scores of $10$, $5$, or $0$ for each question. This provides a clear visualization of how the analyzed websites perform in terms of privacy policies and user control, as evaluated by PrivacyCheck.

Most websites performed well in protecting SSN, with those that did not request SSN receiving full marks. Similarly, most sites demonstrated compliance with the COPPA and did not engage in location tracking. However, the subcategory concerning law enforcement cooperation without user notification scored particularly low, suggesting that many websites are willing to share PII with law enforcement without requiring legal documentation. The lack of data breach notification and data encryption policies is concerning. These are critical elements of user protection, yet none of the websites clearly stated they would notify users in case of a breach, and the low average score for data encryption suggests many sites either do not encrypt data or do not mention encryption in their privacy policies.

\begin{table}[tbp]
\begin{center}
\caption{Privacy policy subcategories under User Control and Policy Framework categories as obtained from PrivacyCheck, along with the mean scores in range $0-10$}
\label{table:privacy_scores}
\begin{tabular}{p{2cm}p{3.5cm}p{1cm}}
\hline
\textbf{Category}     & \textbf{Subcategory}                & \textbf{Mean}        \\ \hline   
User Control & Email Address Security      & $5.06$  \\
             & PII Security                & $6.17$  \\
             & SSN Security                & $9.29$  \\
             & Targeted Advertising        & $3.96$  \\
             & Location Tracking           & $7.14$  \\
             & COPPA Complicance           & $7.34$  \\
             & Law Enforcement             & $2.01$  \\
             & Privacy Policy Opt-Out      & $3.96$  \\
             & Data Control                & $4.74$  \\
             & Data Aggregation            & $4.42$  \\ \hline 
Policy         & Between Site Transfer       & $6.97$  \\
Framework             & Company Location            & $7.24 $ \\
             & Right to be Forgotten       & $6.58$  \\
             & Data Retention Notification & $7.11$  \\
             & Reject Usage of PII         & $8.29$  \\
             & Under 16 Protection         & $6.18$   \\
             & Data Encryption             & $3.68$  \\
             & Data Processing Consent     & $7.37$   \\
             & Data Protection Principles  & $8.16$  \\
             & Breach Notification         & $0$     \\ \hline
\end{tabular}
\end{center}
\end{table}

\textbf{Cookies Analysis Results (RQ2):}
On average, each website in our study deployed around $38.5$ cookies, with the highest number reaching $115$ on a single site. A significant $31.17\%$ of these cookies were categorized under the ``other" category, indicating that their purposes were unclear or not adequately communicated to users. Analytical cookies were the most frequently used, making up $19.47\%$ of the total, while necessary and functional cookies were less common, representing $8.42\%$ and $8.2\%$, respectively. The high proportion of cookies classified as ``other" and the low score for targeted advertising transparency ($3.96$) in the privacy policy analysis suggest that many websites fail to clearly communicate their cookie usage to users. This lack of transparency raised concerns about user awareness regarding how their data is being tracked and utilized.

\begin{table}[tb]
\begin{center}
\caption{Manual evaluation categories with percentage of complying websites}
\label{table:manual}
\begin{tabular}{p{2cm}p{3.25cm}p{1.25cm}}
\hline
\textbf{Category}            & \textbf{Subcategory}                & \textbf{\% Score} \\ \hline
Website Security    & Digital Certificate         & $100$           \\
                    & HTTPS (SSL)                 & $100$           \\ \hline
Authentication      & Basic authentication        & $96.55$         \\
                    & Two factor authentication   & $0$             \\
                    & Allows guest checkout       & $68.97$         \\ \hline
Payment Security    & Secure Payment              & $89.53$         \\
                    & Electronic Signatures       & $83.33$         \\
                    & Association with FI         & $38.46$         \\ \hline
Input Validation    & Address Verification System & $86.36$         \\
                    & Card Verification           & $100$           \\ \hline
Customer Confidence & Customer Reviews/Rating     & $59.30$         \\ \hline
\end{tabular}
\end{center}
\end{table}

\subsection{Manual Analysis (RQ3)}
Table~\ref{table:manual} shows the average percentage score of manual category criteria, aggregated over all the websites. We note that only around $59\%$ of the websites from the sample provided a way for customers to leave feedback on the products and provide rating. 

\textbf{Authentication Analysis:}
Our examination of authentication mechanisms revealed that while most websites implemented at least basic authentication (username and password), a small percentage (around $3\%$) did not. None of the websites mandated MFA, although some may have offered it as an option. Approximately $69\%$ of the websites allowed guest checkout, providing users the convenience of making purchases without creating an account, yet this also raises concerns about identity verification.

\textbf{Payment Security Analysis:}
We assessed how websites communicated their payment security measures to users. Around $90\%$ of the websites indicated that secure payment methods were in place, and $83\%$ offered electronic signatures. However, only $38\%$ of the websites mentioned any association with a well-known financial institution or bank, which could be a critical factor in establishing user trust.

\textbf{Input Validation Analysis:}
Effective input validation is essential to prevent fraudulent transactions and ensure data accuracy. Our analysis found that all websites validated credit or debit card numbers. However, about $14\%$ of the websites failed to validate shipping or billing addresses, which could pose risks to both users and the websites.

\textbf{Website Security Analysis:}
We focused on easily visible security features, such as digital certificates and HTTPS (SSL) connections. As expected, all $90$ websites had valid digital certificates and provided secure HTTPS connections at least during the payment process, which aligns with standard web security practices. However, one site only secured its store page with HTTPS, leaving related pages unprotected.

\section{Implications} \label{implication}

\subsection{Simplifying Policies \& Transparency in Cookie Usage} Our findings highlight significant user control issues, with the average control score at $53.4\%$, indicating that users often lack clear information about how their data is handled. To address this, we recommend that e-commerce platforms simplify privacy policies using plain language and interactive privacy dashboards~(\cite{R103}). These dashboards can visually present data handling practices, making them more transparent and easier for users to understand. Additionally, it's essential for platforms to clearly communicate the purpose and function of cookies, particularly those categorized under ``other." This clear communication will enable users to make informed decisions, avoiding broad consent options and dark patterns~(\cite{grassl2021dark}). Platforms should provide users with granular choices about cookies and allow easy adjustments to these preferences at any time~(\cite{habib2022okay}).

\subsection{Risk Communication and Management} The absence of clear breach notification protocols, with all websites scoring $0$ in this area, is a serious concern. E-commerce platforms must develop robust risk communication strategies, ensuring that users receive prompt and transparent notifications in the event of a data breach. These notifications should clearly explain the nature of the breach, its potential impact, and the steps users should take to protect themselves~(\cite{ko2004cross}). Additionally, we recommend that platforms adopt Risk-based Authentication (RBA) as a dynamic layer of security. RBA can evolve with user behavior, incorporating methods like behavioral biometrics and device fingerprinting to enhance security and user experience~(\cite{wiefling2020more}).

\subsection{Account Remediation Protocols} Our analysis revealed a significant gap: none of the examined websites had established account remediation protocols in place. This absence of clear guidelines for notifying users in case of a security breach or compromised account integrity is concerning. E-commerce platforms must prioritize the development of comprehensive account remediation processes. These protocols should include immediate breach notifications, clear instructions for securing accounts, and accessible channels for users to seek help. By implementing these measures, platforms can better protect their users, minimize damage from breaches, and demonstrate a strong commitment to data security~(\cite{neil2021investigating,markert2023transcontinental}).

\subsection{Enhancing Website Security Practices} While all websites in the study adhered to basic security practices like SSL and HTTPS, more advanced measures are necessary. None of the websites mandated multi-factor authentication (MFA), a critical security feature. E-commerce platforms should implement MFA as a standard requirement, particularly for transactions involving sensitive information. Furthermore, only $38\%$ of websites clearly communicated their associations with well-established Financial Institutions. Platforms should be transparent about their affiliations with reputable financial institutions, as this transparency can significantly enhance user trust in the platform's security. Secure payment gateways and proper input validation of addresses and credit card information are also crucial to protecting both users and platforms from fraud and security threats.

\subsection{End-User Empowerment} Empowering end-users is essential for enhancing security in e-commerce. We found that tools like PrivacyCheck and Cookieserve can play a crucial role in helping users assess the privacy and security practices of websites~(\cite{gonzalez2009measurement}). E-commerce platforms should optimize these tools for ease of use, making them more accessible to the average user. Additionally, platforms can offer users a checklist for evaluating site integrity, covering aspects such as SSL certifications, secure payment gateways, and customer reviews. By partnering with cybersecurity experts, platforms can also provide educational resources, such as workshops and webinars, to help users navigate online shopping safely. Finally, establishing an open feedback loop where users can report suspicious activities or share their experiences can create a community-driven approach to security, enhancing the overall safety of the platform.

\section{Future Work and Limitations}
\label{future}
Our study provides significant insights into the privacy and security practices of e-commerce websites, yet there are limitations. The automated tools used, such as PrivacyCheck and Cookieserve, while effective, might not fully capture the complexity and nuances of privacy policies and cookie usage across diverse platforms. Additionally, the absence of direct user interaction restricts our understanding of how users perceive and respond to the privacy and security measures employed by these websites. In the future extension of this work, we plan to address these limitations by extending our dual-technique analysis through a naturalistic experiment, which will focus on the final phase of online transactions, particularly assessing post-payment verification processes such as email and SMS confirmations to understand the type and amount of user data shared in these communications. Furthermore, we aim to broaden our research to include transcontinental websites, which will allow for a more comprehensive evaluation of GDPR compliance and its interaction with other international regulations, providing a global perspective on e-commerce privacy and security practices.

\section{Conclusions} \label{conclusion}
In today's digital economy, securing e-commerce websites is crucial, particularly for platforms frequented by younger users. Our analysis of $90$ US-based shopping websites employed a dual-technique approach, utilizing automated tools like PrivacyCheck and Cookieserve alongside detailed manual evaluations. The manual evaluation focused on areas such as payment security, authentication, and input validation. We identified significant issues, including a lack of opt-out options, with only $3.96\%$ of websites offering clear data control mechanisms. Notably, all websites provided digital certificates and SSL, yet none mandated MFA. The prevalence of uncategorized ($31.17\%$) and advertisement ($24.04\%$) cookies further raises concerns. To address these, we recommend simplifying privacy policies, enforcing robust account remediation protocols, and enhancing transparency in risk communication to ensure compliance with regulations like CCPA and COPPA, thereby fostering greater trust and safer online experiences for all users.

\addtolength{\textheight}{-.2cm} 

\bibliographystyle{apalike}
\bibliography{HICSS2025.bib}


\end{document}